\documentclass[11pt,twoside]{article}
\usepackage{./asp2014}
\usepackage{xspace}

\newcommand{\kms}{\ensuremath{\rm{km\,s}^{-1}}\xspace}
\newcommand{\alphaCO}{\ensuremath{\alpha_{\rm{CO}}}\xspace}

\newcommand{\Mstar}{\ensuremath{M_{\rm{star}}}\xspace}
\newcommand{\Mgas}{\ensuremath{M_{\rm{gas}}}\xspace}

\newcommand{\Mout}{\ensuremath{M_{\rm{out}}}\xspace}
\newcommand{\Msol}{\ensuremath{\rm{M}_\odot}\xspace}

\newcommand{\tdep}{\ensuremath{t_{\rm{dep}}}\xspace}

\newcommand{\oiii}{[O{\scriptsize III}]\xspace}
\newcommand{\halpha}{\ensuremath{\rm{H}\alpha}\xspace}

\newcommand{\arc}{\ensuremath{''}\xspace}

\aspSuppressVolSlug
\resetcounters

\begin{document}

\title{Characterizing Feedback Through Molecular Outflows Across Cosmic Time}

\author{Justin~Spilker$^1$ and Kristina~Nyland$^2$}
\affil{$^1$University of Texas at Austin, Austin, TX, USA; \email{spilkerj@gmail.com}}
\affil{$^2$National Radio Astronomy Observatory, Charlottesville, VA, USA; \email{knyland@nrao.edu}}

\paperauthor{Justin~Spilker}{spilkerj@gmail.com}{0000-0003-3256-5615}{University of Texas at Austin}{McDonald Observatory \& Department of Astronomy}{Austin}{TX}{78712}{USA}
\paperauthor{Kristina~Nyland}{knyland@nrao.edu}{0000-0003-1991-370X}{National Radio Astronomy Observatory}{}{Charlottesville}{VA}{22903}{USA}

\begin{abstract}
Galactic winds are ubiquitously observed in galaxies both locally and in the high-redshift Universe. While these winds span many orders of magnitude in both temperature and density, observations of nearby galaxies show that the cold molecular phase tends to dominate both the mass and momentum carried. The capabilities of the ngVLA for the study of molecular outflows at low redshift are described elsewhere in this Volume; here we focus on the ability of the ngVLA to detect and image such outflows in the high-redshift Universe via deep observations of low-J transitions of the CO molecule. The ngVLA is capable of detecting molecular outflows from typical galaxies on the star-forming sequence with $\log \Mstar/\Msol \gtrsim 10.5$ to $z\sim3$, and galaxies with higher star formation rates to beyond $z\sim4$. The ngVLA will enable an understanding of the feedback processes that shape galaxies throughout the epoch of galaxy assembly when the bulk of the stars in the Universe were formed. While the emission associated with outflows is faint in comparison to the emission from the galaxy, deep observations are also required for high-resolution dynamical studies, allowing for the routine simultaneous detection and imaging of the outflows.
\end{abstract}

\section{Galaxy Feedback Through Cosmic Time}

One of the most important realizations of the last fifteen years is the vital role that feedback processes must play in the evolution of galaxies.  The energetics of feedback processes are critical unknowns in theoretical models of galaxy evolution. Energy and momentum from active galactic nuclei (AGN), Type~II supernovae, and/or radiation pressure from massive stars are input in ad hoc manners below the resolution limit of simulations. These feedback processes play critical roles in the evolution of galaxies by regulating the growth of stellar mass in relation to dark matter halos, establishing a connection between stellar velocity dispersion and supermassive black hole mass, and enriching the circumgalactic medium with metals out to large radii. 

At the extreme end, high-resolution hydrodynamical simulations currently favor dramatic processes such as starbursts triggered by gas-rich mergers or disk instabilities to deplete, expel, or heat the gas supply and prevent star formation \citep[e.g.,][]{hopkins08,ceverino15,wellons15}. Feedback also remains critical even for less extreme galaxies -- state-of-the-art treatments of feedback yield `bursty' star formation histories, with star formation rates (SFRs) that vary by an order of magnitude or more on timescales of $\sim10-100$\,Myr \citep[e.g.,][]{hopkins14}. This burstiness is related to the ability of feedback to quickly and efficiently disrupt star-forming gas until the gas can cool and resume forming stars. 

Observational evidence for the role of feedback has grown substantially in recent years thanks to the discovery of ubiquitous gas outflows in local and moderate-redshift galaxies. Powerful outflows have been observed in both ionized \citep[e.g.,][]{forsterschreiber14,genzel14} and neutral gas \citep[e.g.,][]{shapley03,rupke05}. Critically, however, the discovery of ubiquitous massive, energetic \textit{molecular} winds in low-redshift star-forming galaxies, (Ultra)Luminous Infrared Galaxies ((U)LIRGs), and AGN host galaxies has provided evidence that feedback can at least temporarily suppress star formation by acting \textit{directly} on the fuel from which stars form \citep[e.g.,][]{feruglio10,veilleux13,cicone14,leroy15}. 

Because the spectral signatures of outflows are very faint, very few detections of molecular outflows are known at higher redshifts \citep[e.g.,][]{cicone15,spilker18b}. Nevertheless, the high-redshift regime ($1\lesssim z \lesssim 4$) represents a critical epoch in the history of the Universe: the cosmic SFR density peaked at $z\sim2-3$, and most of the stars in the Universe were formed prior to $z=1$ \citep[e.g.,][]{madau14}. Understanding the role of feedback in regulating the growth of galaxies at this epoch is thus of great importance. The prospects for the ngVLA to characterize galactic feedback processes in low-redshift galaxies are described elsewhere in this volume \citep[i.e.,][]{bolatto18ngvla}, and we refer the reader to that chapter for additional discussion of the importance of feedback processes and key outstanding questions to be addressed by observations of molecular outflows. Here, we focus on the capabilities of the ngVLA to detect and image molecular outflows across the peak of cosmic star formation.

\section{High-Redshift Discovery Space for the ngVLA}

As currently envisioned, the ngVLA will have $\approx10\times$ the sensitivity of either the existing Karl G. Janksy VLA or ALMA. This large increase in sensitivity, especially in the 3-4\,mm atmospheric window, is critical for the detection of molecular outflows at moderate redshift. Molecular outflows have been primarily observed through far-infrared absorption lines of OH and, of relevance for the ngVLA, low-J CO lines. In the case of CO, the signature of winds is evident as additional emission at high relative velocities to the galaxy systemic velocity. If the spatial resolution of the observations is low, outflows can be identified from the integrated CO spectrum alone -- in order to reasonably expect to detect the faint outflowing gas, the galaxy itself will be very significantly detected with a very well-characterized line profile. If the spatial resolution is high, the outflow can additionally be distinguished because it is not expected to follow the predominant galactic kinematics (e.g., the outflow is likely to be launched from the minor axis perpendicular to a galaxy's rotation curve).

\subsection{Molecular Outflow Detection Limits}

To demonstrate the capability of the ngVLA to detect molecular winds at high redshift, we perform a simple simulation. As a template, we use the outflows observed in low-redshift U/LIRGs in CO by \citet{cicone14}. These galaxies have typical SFRs $\sim$100\,\Msol/yr and outflow masses $\Mout \sim \mathrm{few} \times 10^8$\,\Msol. The molecular mass contained in the outflows typically comprises 1--10\% of the total galaxy molecular gas mass. Because CO(1--0) is not accessible for all redshifts, we also simulate CO(2--1) and (3--2) emission, assuming line brightness temperature ratios to CO(1--0) of $r_{21} = 0.8$ and $r_{31} = 0.5$.  We note that many scaling relations between outflow properties and galaxy properties are poorly-constrained even at low redshift and are completely unknown at high redshifts; this exercise is merely meant to be illustrative.

We use the current best estimates for the array performance, assuming the array sensitivity for 1\arc imaging in order to avoid resolving out the outflows as much as possible. This effectively corresponds to considering only the antennas that comprise the `core' of the array on baselines $\lesssim$1.3\,km. We assume a 10\,hr integration time and 50\,\kms channelization of the data. The result is shown in Figure~\ref{fig:spec} for three hypothetical galaxies at redshifts $z = $0.5, 1.5, and 2.5, with the outflow observed in CO(1--0), CO(2--1), and CO(3--2), respectively. In each case, high-velocity emission corresponding to the molecular wind is easily detected (the emission from the galaxy itself is also detected at a signal-to-noise of many tens per channel).  

\articlefigure[width=\textwidth]{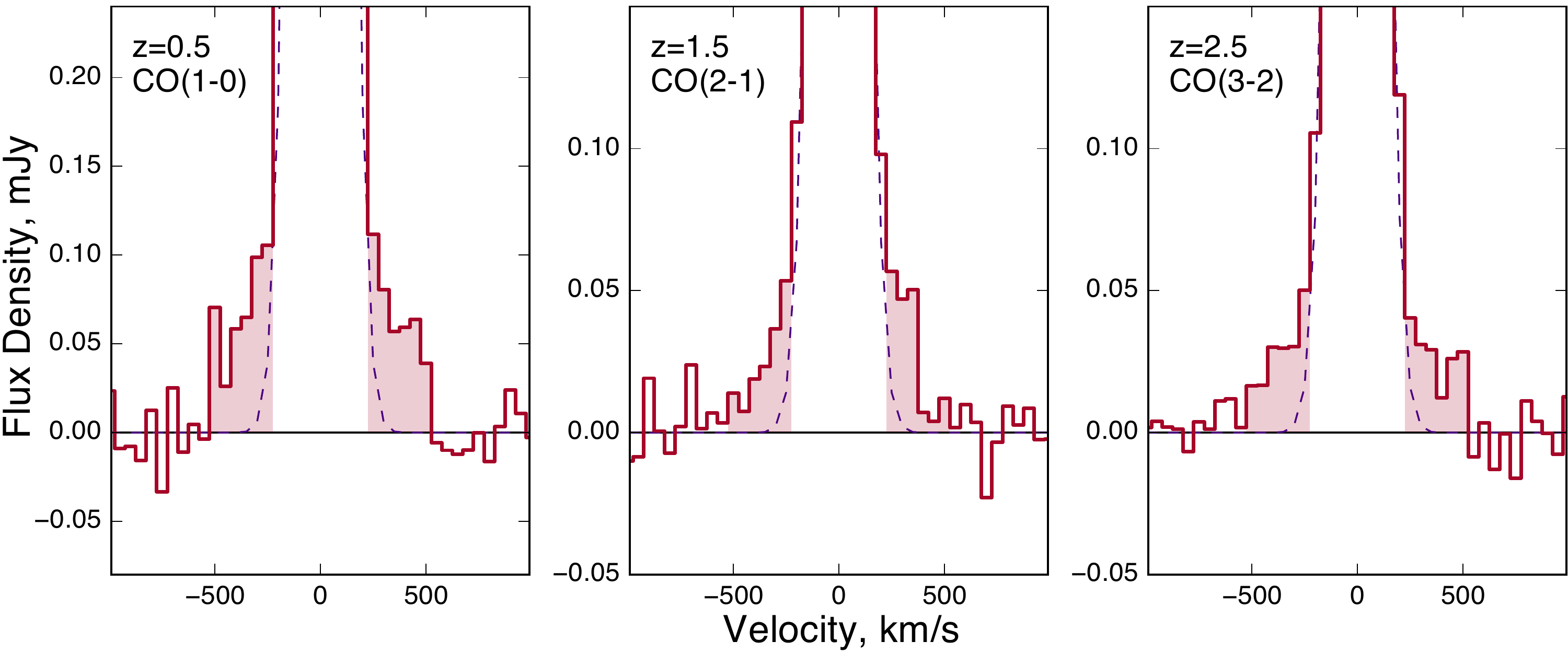}{fig:spec}{Three example CO spectra of galaxies at high redshift, showing the capability for the ngVLA to detect molecular outflows via high-velocity wings in the spectra that are distinct from the emission from the galaxy (illustrated as Gaussian fits to the spectra core, which extend well above the upper y-axis limits). We use outflows observed in low-redshift U/LIRGs (SFR$\approx$100\,\Msol/yr, $\Mout \sim$ few $\times 10^8$\,\Msol; \citealt{cicone14}) as templates to create these spectra, assume a 10\,hr integration, and use the expected ngVLA sensitivity for 1\arc imaging. In each case high-velocity emission is well-detected.}

Detections of similar outflows using existing facilities would be virtually impossible: the current VLA lacks receivers that can access the 3-4\,mm atmospheric window where the CO emission from outflows is brightest, and reaching equivalent depths with ALMA would require of order 300-500 hours of observation. Only the order-of-magnitude increase in sensitivity afforded by the ngVLA makes the detection of outflows such as these possible in the distant Universe.

In Figure~\ref{fig:mout} we show the 5$\sigma$ detection limits for molecular outflows as a function of redshift and targeted CO transition. We assume a CO-H$_2$ conversion factor $\alphaCO = 0.8$\,\Msol\,(K\,\kms\,pc$^2$)$^{-1}$ and a 500\,\kms outflow velocity width. For all redshifts $z>1$, more sensitive mass limits can be reached using CO(2--1) or CO(3--2), even if the CO excitation is low (little is known about the CO excitation in existing detected winds). Because the outflow mass \Mout is a rather abstract quantity, we label the right-hand axis of Figure~\ref{fig:mout} in terms of the approximate corresponding galaxy SFR, assuming a linear correlation between \Mout and SFR. We assume the outflow mass constitutes 3\% of the total galaxy gas mass, and convert to an SFR assuming a gas depletion time $\tdep = \Mgas/\mathrm{SFR} = 150$\,Myr. This `conversion' is obviously extremely simplistic and uncertain, and the true scaling relations between outflow properties and galaxy properties would be a key outcome of a future ngVLA campaign. We merely aim to place the detectable outflows in terms more familiar to the extragalactic community using known characteristics of low-redshift outflows.

Figure~\ref{fig:mout} demonstrates that outflows similar to those in the nearby universe are detectable by the ngVLA out to high redshift. In particular, the ngVLA can detect outflows even from `typical' massive galaxies -- we illustrate the SFR of galaxies on the star-forming (``main'') sequence for galaxies with $\log \Mstar/\Msol = 10.5$ and 11 as a grey shaded region in Figure~\ref{fig:mout} \citep{speagle14}. The ngVLA will place outflows from normal galaxies within reach, and should comparatively easily detect outflows from more highly star-forming objects. At the very highest redshifts, $z>4$, only winds from the most highly star-forming objects are likely to be accessible (e.g., the population of high-SFR dusty star-forming galaxies; \citealt{spilker18b}), a consequence of the inexorable cosmological dimming of spectral features.

\articlefigure[width=\textwidth]{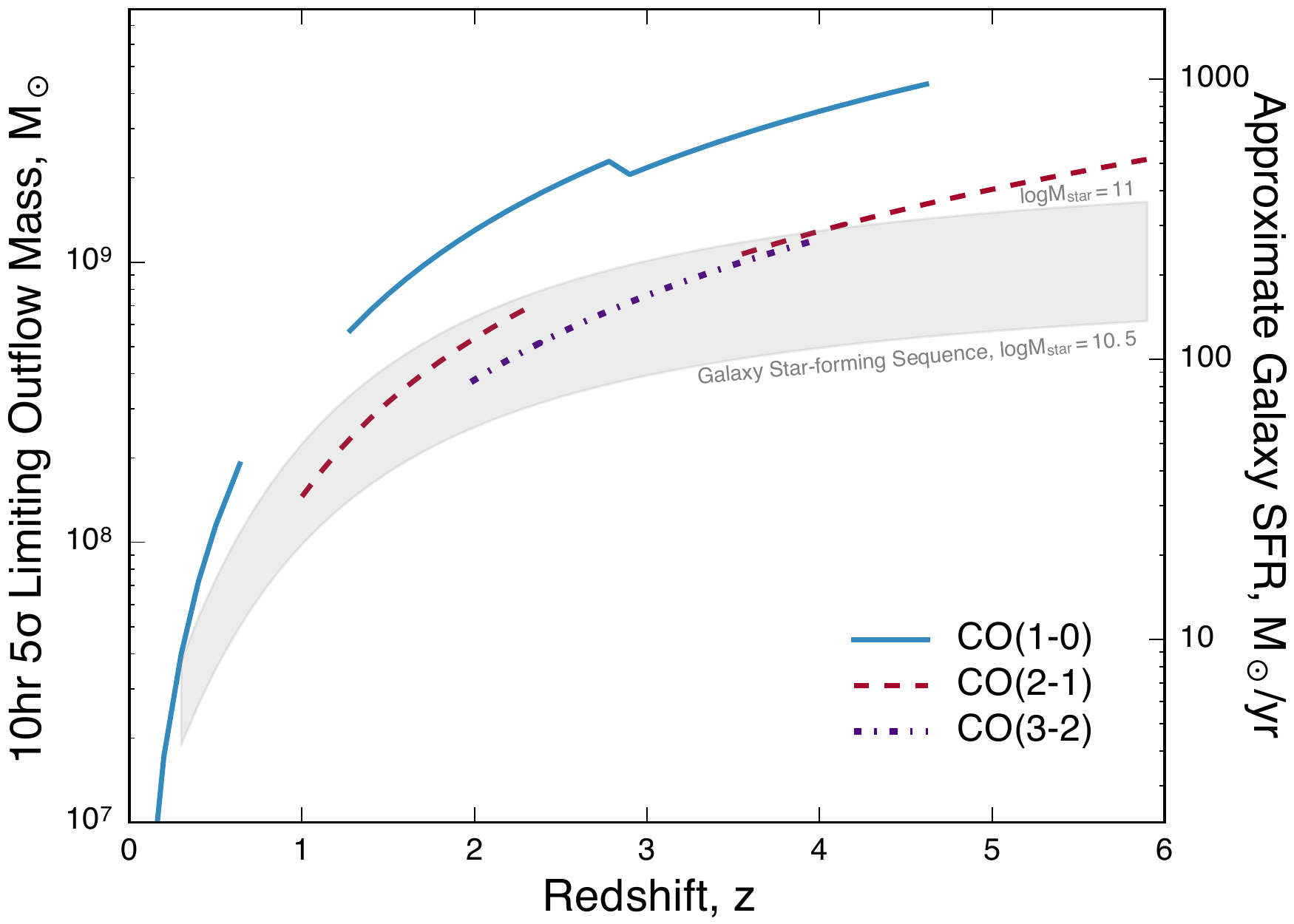}{fig:mout}{The approximate 5$\sigma$ outflow mass detection limits as a function of redshift. Gaps in the redshift coverage for a given CO transition are due to the telluric oxygen complex near 60\,GHz. Because the outflow mass \Mout is a rather abstract quantity, the right-hand axis shows approximate corresponding galaxy SFRs (see text for details); this axis is merely intended to place the detectable outflows in more familiar context. We also show the SFR corresponding to the galaxy star-forming sequence for galaxies with stellar masses of $\log \Mstar/\Msol = 10.5$ and 11 as a grey shaded region \citep{speagle14}. The ngVLA will detect molecular outflows from typical massive galaxies on the star-forming sequence out to $z\sim4$ as well as from starburst galaxies well above the star-forming sequence.}

\subsection{Prospects for High-Resolution Imaging of Outflows}

The previous subsection demonstrated the detectability of molecular winds assuming fairly low-resolution 1\arc imaging. Such observations effectively only include the core of the ngVLA, with $\approx$44\% of the total collecting area on baselines $<$1.3\,km. The second `tier' of longer baselines extends to $\approx$30\,km and adds a further $\approx$35\% of the total collecting area. High-resolution observations are thus possible, at the expense of spreading the fixed total source flux over multiple resolution elements.

Precious little is known about the structure and geometry of molecular winds even at low redshifts, and even less is known in the distant universe. The ngVLA will have the capability to directly image molecular winds at high redshift, if the CO emission is sufficiently bright and/or concentrated into bright clumps. At 0.1\arc resolution, for example, the ngVLA will reach sensitivities $\sim$20\% better than at 1\arc resolution, but this is insufficient to compensate for the 100$\times$ smaller beam area. Thus, the surface brightness sensitivity of the ngVLA will necessarily limit the imaging of outflows to only the brightest, most massive clumps contained within the winds. Current simulations predict extremely clumpy outflows due to hydrodynamical instabilities \citep[e.g.,][]{richings18,schneider18}, so the detection of wind substructure on scales accessible to the ngVLA may yet prove feasible.

We note also that one of the key science drivers of the ngVLA is the ability to image low-excitation CO emission at high resolution and sensitivity \citep[see, for example,][in this Volume]{carilli18ngvla}. Such observations face challenges very similar to those we identify here -- namely, faint CO emission requires significant integration times to detect at high resolution. Carilli \& Shao present simulations that assume 30\,hr observations in order to characterize the dynamical properties of an M51-like galaxy at redshifts $z=0.5-4.2$. Such observations would also be sensitive to subkiloparsec-scale clumps of molecular material in a molecular wind with a mass limit $\approx2\times$ below the thresholds shown in Figure~\ref{fig:mout}. Thus, perhaps the most promising pathway to detailed imaging outflows by the ngVLA will be to simply make use of observations with a primary goal of characterizing the detailed dynamics of galaxies.

\section{Towards a Multi-phase Understanding of Outflows: Complementarity with Future Facilities}

Given its fundamental importance in the evolution of galaxies, feedback will continue to be a focus of study for the foreseeable future. The ngVLA will have the ability to detect, and in some cases image at high resolution, the molecular phase of outflows at high redshift. Outflows are however complex, spanning many orders of magnitude in temperature and density, and so different phases of the winds are observable across the electromagnetic spectrum \citep[e.g.,][]{leroy15}. Again, we refer the reader elsewhere in this volume for a more detailed discussion of the capabilities of envisioned future facilities in the study of outflows \citep{bolatto18ngvla}. Here we merely highlight a few facilities in the context of high-redshift work.

By the time the ngVLA enters operations, 30\,m-class ground-based optical telescopes will have made great advances studying the warm ionized phase of outflows through observations of \halpha and \oiii. The Giant Magellan Telescope, Thirty-Meter Telescope, and European Extremely Large Telescope all have plans for near-infrared integral field units, allowing studies of ionized outflows at matched or better resolution than will be feasible with the ngVLA to $z\sim2.5$. Meanwhile, the \textit{James Webb Space Telescope} will have observed outflows in a similar fashion to even higher redshift at somewhat worse spatial and spectral resolution. The winds detected by \textit{JWST} are likely to become well-studied `template' outflows, and the galaxies hosting such winds will be natural early targets for ngVLA observations of the molecular phase. These same facilities will also likely constrain the neutral atomic component of outflows through the detection of low-ionization metal absorption lines (e.g., Na~I, Mg~II), although these lines are far fainter than the bright optical strong lines.

Finally, the proposed NASA flagship \textit{Origins Space Telescope} (OST) would allow for comprehensive views of the large majority of outflowing gas to very high redshift ($z\sim5$). OST would provide a thousandfold increase in spectral sensitivity and mapping speed compared to previous far-infrared facilities, allowing the detection of outflows in higher-redshift, lower-mass, lower-SFR galaxies. While the spatial resolution of OST would necessarily be much lower than that of the ngVLA, OST will observe molecular, neutral, and warm ionized phases of outflows in thousands of star-forming galaxies through the detection of P Cygni profiles in far-infrared lines of OH and high-velocity wings of fine structure lines. Molecular winds detected by OST would be prime targets for ngVLA followup. The ngVLA will allow for both high-resolution imaging of the outflowing gas, and provide better estimates of the outflow masses and mass loss rates because CO observations are not limited to only the line-of-sight component of the wind as in absorption studies.

\acknowledgements J.S. thanks the University of Texas at Austin for support through a Harlan J. Smith Fellowship.

\bibliography{ngVLA_highz_outflows.bbl}

\end{document}